# TCLOUD: Challenges and Best Practices for Cloud Computing

Sultan Ullah, Zheng Xuefeng, Zhou Feng, Zhao Haichun

*School of Computer and Communication Engineering, University of Science and Technology Beijing, China*

sultan.ustb@yahoo.com, zxf@ies.ustb.edu.cn, ustbzhou@126.com, cieczhaohc@163.com

**Abstract**

*Cloud computing has achieved an unbelievable adoption response rate but still its infancy stage is not over. It is an emerging paradigm and amazingly gaining popularity. The size of the market shared of the applications provided by cloud computing is still not much behind the expectations. It provides the organizations with great potential to minimize the cost and maximizes the overall operating effectiveness of computing required by an organization. Despite its growing popularity, still it is faced with security, privacy, and portability issues, which in one or the other way create hurdles in the fast acceptance of this new technology for the computing community. This paper provides a concise all around analysis of the challenges faced by cloud computing community and also presents the solutions available to these challenges.*

*Index Terms: Security Challenges, Best Practices, Cloud Security, Attack on Cloud, Privacy Protection.*

## 1. Introduction

Cloud Computing is known to be the set of all those inter related components that work together in order to make resource sharing possible. It has evolved through a great level of implementations. It provides the user the convenience of moving data and also scalability of infrastructure, application development platforms and valuable business application platforms [1].

Presently cloud computing is considered to be one of the most talk about topics in computing world. Though, the term "Cloud Computing" does not represents a set of whole new technologies, but rather it present the mechanism to combined the existing technologies so that it enable the customer to create new computing services and business model. It is evident that at the beginning of computing era, most of the time there would be a single computer, which would be present at some distant data centre and the resources would be shared by many users.

As the internet is considered to be the set of clouds, thus it can be defined as to use internet for the creation, dissemination of technology enabled services for the users of computing world [1][2].

The cloud user avail the services provided by the cloud provider through internet anytime and from anywhere across the globe, without bothering to know the physical or technical management of the cloud resources.

It is apparent that as with many new product/ technologies and services provided by the computing industry, in cloud computing information and data security are the most important issues and it is very much discussed. These issues are closely observed critically than with the offering of the cloud services which are around for some time. The potential user of the cloud services have great concern about data/ information security which is the main hurdle in the deployment of the cloud on a large scale. It is necessary for cloud service provider to developed trust among its customer which is the basic requirement to take advantages from the offerings that are provided by the cloud [3].

Despite the fact that the demand is on increase day by day around the globe for the services from the cloud, but almost every study and survey point out some major concerns of the cloud users which dispirit the users away from using Cloud Computing services. One of the main hurdles which are mentioned frequently is a lack of faith in the security of the services provided.

The world is attracted towards the cloud due to its many advantages but as every picture has two sides, the cloud has various problems to overcome as well. Due to the increase of interest in the cloud services around the globe, the hackers also found it to be a new and best play ground for their activities. The activities from the hackers will not only affect the users but it will also have a great affect on the cloud service provider.

The rest of the paper is organized as; literature review is presented in 2, cloud challenges and best practices are presented in 3, and the paper is concluded in 4.

## 2. Literature Review

The customers of the cloud services have shown great concern on the security and the way to access the cloud computing environment. It is important for a Cloud Service Provider (CSP) to give surety to its user that the information stored on the cloud site is secure, safe and it cannot be accessible by unauthorized personnel. A cloud security framework which will trace the movement and processing of information which are available for processing on the cloud, the system required that there should be a security capture device on the cloud. The functionality of the device will be to ensure the cloud user that their information is safe from any security threat or attack [4].

The proposed tool for data centric security named Declarative Secure Distributed System (DS2). The DS2 provides a detail system forensics and analysis, well-organized end-to-end authentication and verification of data, secure query processing, and seamless integration of declarative access control policies [5].

The threat of illegitimate access in cloud computing is high due to the non availability of a tough access control mechanism. When the access control policies / mechanisms are not appropriate then there is a strong possibility that even classified data may be accessed illegally. In a cloud environment where multiple users will use different protocols to access the services from a single entity, make it very difficult for the cloud service provider to provide an interoperable identification and authentication mechanism [6].

The cloud computing environment is facing many surety threats. The "Wrapper Attack" is one of these threats, in which some malicious codes is wrapped in XML signature and inject that to the XML code which is essential in cloud computing for the sharing of resources [7].

The cloud present pay – as – you go model, and one of the other common issues which is normally known as the flooding or Denial of Service (DoS) attacks. If the cloud server is occupied by the requests sent the DoS attacker, then the server will denied its services to other users which are requested as the server on cloud is busy due to serving of the DoS requests. The situation will be even more severe if the hacker gets access to more nearby server machines to send the flood requests, which is called Distributed Denial of Service (DDoS) [8].

A framework of security system that will provide secure authentication mechanism, privacy and security of secret documents/files, to control unauthorized access, and provide track mechanism to resolve disputes of data, is presented [9].

The rate of cyber crime / hacker attacks on cloud computing environment is on rise, and it consider to be the tempting play land for the hackers [10]. A cloud is the combination of several heterogeneous entities and security of such an environment is very difficult. Numerous privacy issues like retention and destruction of data is also there [11]. The cross border transition of data also presents another security issue as there is a difference in policies among different countries [12].

IP Spoofing is another very serious security challenge for the cloud computing. The Virtual Machines (VM) are assigned an IP address and the hackers try to have these IP addresses of the VMs. In some cases it results in access to the confidential of the users [8].

It is a fact that the user will be no more in control of its data once it is outsourced to the cloud environment. Apparently it is believed that no one is in a position to say anything about the data, where it will be stored, where it will be processed, and who has the right to access and who cannot.

## 3. Challenges and Best Practices for Cloud Computing

Presently at all levels of the cloud computing platform, whether it is the network layer, web application layer or host layer there exist a corresponding security threat. Such types of security issues have been studies extensively by the professionals in the field of information security. Cloud computing security need to be focus on the analysis and providing solutions for cloud computing service computing model, dynamic virtualization management, and multi-tenant shared operational mode for data security and privacy protection [13].

The cloud customers (company) trust the Cloud Service Provider (CSP) when it outsource its sensitive data or run applications on a remote site of the CSP. The computer system will be vulnerable to inside threats, although by having the most sophisticated firewalls and computer security mechanism. Due to the dereliction in duty by internal staff, hacker attacks and system failures lead to a variety of risks to security mechanisms such as the data lose, cloud service providers assures users that their data is safe and protected.

### 3.1. Authentication and Authorization

De Refrain unauthorized users to avail the services or to get access to confidential data stored on the Cloud is of great importance. The mechanism to do so is to provide the authorized user with a systematic mechanism of authentication and authorization by the Cloud CSP. As a matter of fact that the customer needs its data to be accessible by its user only, but the system administrator will also have the full access to the data, which is responsible for looking after the cloud environment. It is of great importance to create and maintain a firm confidence level between cloud service provider and its customer [13].

The universal method to preserve authorized access to cloud computing sites by using the web browsers is the use of role based access and password protection. Instead of having only user name password as a method of Authentication and authorization, we should also provide some additional authentication factor as well [13].

### 3.2. Dynamic Virtualization Management

In a typical clouds computing service platform, the resources are virtually available on pay-as-you go mode. These virtual resources are presented by limited actual physical resources. In a multitenant shared resource cloud computing environment in which the customer shared the same amount of physical resources but virtually it looks to be separated from one another physically as well, which is not the case. So cloud platform virtualization presents security vulnerabilities, in short the user data can be accessed by other user some time [14] [15] [16].

The technique used for virtualization which allows multiple guest operating systems to run concurrently is known to be a Hypervisor. Normally it is embedded in the kernel of the host operating system or hardware infrastructure. It provides additional security tools e.g Intrusion detection Systems, but it is still vulnerable. If by any means hypervisor crashed or hacker gets control over it then all VMs are on the attackers' control. Although to take control over the hypervisor is difficult but not impossible. The easiest way to safe guard against the above mentioned threat, it is recommended to have an updated version of the virtualization product.

### 3.3. Portability and Interoperability

Portability refers to the ability to move application and its data from one cloud to another. Portability could be achieved by removing dependencies on the underlying atmosphere. A portable components (application, data) could be moved and reused regardless of the provider, platform, operating system, location, storage etc without being modified e.g. if the old cloud environment is Windows and new cloud environment is Linux then an application running on old cloud would be able to run on new cloud without being changed is called portability.

The Interoperability and portability present another open research problem for the researcher. Interoperability is the way how different clouds would communicate. It refers to the ability of customers to use the same parameters-management tools, server images etc- with a variety of cloud computing providers and platforms e.g. Amazon and Google are two clouds. Using the same image of Windows from Amazon on Google without any change is called interoperability. This would require Google to understand Amazon language.

The cloud lacks the computing standards to export or import data, storage of elements and process for disaster recovery [17]. To overcome the issue of incompatibility and vendor lock in, the cloud computing community should move towards open standards.

### 3.4. Secure Host and Guest Operating System

It is important to note that a host operating system which is easy to install, use and maintain is required. The host operating system should be up to date and above all secure to maximum extent, because if a hacker got control of the host operating system, then all the guest operating system are also within the control of the hacker. In order to avoid such catastrophe it is advised that host operating system should be kept up to date all the time [18].

A Virtual Private Server (VPS) can be created, modified and deleted by the customer themselves. It is one of the advantages of the virtualization technology that customers select which type of operating system is needed. As all the guest operating systems are available on a single physical machine and having varying level of security, so it easy for a hacker to find vulnerabilities in one the guest operating system.

It is the responsibility of the customer to maintain an up to date operating system and all the required software by them.

### 3.5. Privacy Protection

Privacy is considered to be one of the most talks about issue that exists in all the stages of data life cycle from data generation to data destruction. The challenge is to share data and protect the personal information. Normally all the system needs certain level of privacy, but the most important of all are the system which store financial or health care data.

The growing concerns for the computing community are to control the level of information to reveal and most importantly who can have the access to such information. To be able to provide privacy protection in the cloud environment is to isolate the data which is sensitive from less sensitive data.
In order to solve the problem of privacy, a special encryption scheme is used which is fully homomorphic [19].

The privacy of the data is preserved while using this mechanism. Another framework for the protection of privacy is proposed which is based on the component of information accountability. The information accountability agent have the ability to identify the user, which in way or the other have an access to the information and can also identify the information they use. When there is any improper use of information is detected, the agent will automatically define a set of techniques to hold the users responsible for abuse [20].

### 3.6. Data Confidentiality and Integrity

The storage of data on the cloud is similar to the storage mechanism of the rest of the places, but one thing which should be taken care of is to consider the security of information. The different aspects to ensure security of information are: confidentiality, integrity and timely availability. The problem of confidentiality can be addressed by implementing a sound encryption mechanism. The vendor of the cloud computing services has access to your company's data which hosts VPS, as all the VPS are controlled by a single host operating system and the vendor has full access to that. Therefore all the company's data should be encrypted to protect it against any misuse [18]. In order to take care of the data transfer from the company's terminal to the cloud site and vice versa is also important, and the data stream should be encrypted using SSH – tunneling or VPN.

The data integrity is another challenge, which should also be addressed as data confidentiality. The user outsource / store its huge amount of data on the cloud storage, there is no concrete mechanism to check the reliability of data. The checking of integrity directly on the cloud storage with first downloading it also present another challenge for the computing industry, as a lot of network bandwidth is required in order to download and then upload the data back on to cloud. The traditional ways of ensuring the integrity of data may not be effective in the case of cloud storage.

The problem with data encryption is the management of keys. As a matter of fact the user is not expert enough to take care of their key, and the CSPs will have to maintain enormous keys for the users. A solution to these types of issue is being provided in the form of Key Management Interoperability Protocol (KMIP) [21].

The mathematical model to authenticate the integrity of stored data dynamically is being proposed in [22]. The proposed mechanism of pairing – based provable data integrity (PDI), which not only enables the customer but also a third party verifier to check the integrity of remote data [23].

### 3.7. DoS and DDoS Threat and Responses

The most powerful and dangerous attack on the cloud environment is the Distributed Denial of Service (DDoS) attacks. The CSP's are the most vulnerable to this sort of attack and may shutdown the services they provide. It has a much higher impact on the cloud as oppose to single tenanted structure, because in cloud environment a large number of user share the same resources. To overcome this issue the cloud provider must provide an enhanced Intruder Prevention System (IPS) [24]. But it is observed that in most cases if the DDoS attacks are unidentified and don't have a pre – existing signature, the IPS don't perform the intended functionality.

Another way to protect the cloud against the DoS, DDoS attack is by employing a firewall. It is mandatory for all the CSP's to provide an absolute firewall solution. The simplest way to check the network traffic for DoS, DDoS attack is check the validity of the source IP address. Reverse firewall is another protection option that is provided by the firewall.

### 4. Conclusion

Most of the challenges identified are not novel, they are observed in other field of computing as well, yet the impacts is intensified on cloud computing. Because of the characteristics of the cloud like resource sharing and multi – tenancy, so due to actions of a single tenant can affect all the users of the computing environment, as they share the same interfaces and resources. As a matter of fact, moving towards cloud requires several parameters to be taken care of, but the most important of all is the security.

### 5. References

[1] Sultan Ullah, Zheng Xuefeng, "Cloud Computing: a Prologue", *International Journal of Advanced Research in Computer and Communication Engineering* Vol – 1 No – 1, pp 01 – 04.

[2] Luis M. Vaquero, Luis Rodero-Merino, Juan Caceres, Mike Lindner, "A Break in the Clouds: Towards a Cloud Definition," *ACM SIGCOMM Computer Communication Review*. Vol – 39 No – 1, pp 50 – 55.

[3] Sultan Ullah, Zheng Xuefeng, "Cloud Computing Research Challenges," *IEEE 5th International Conference on BioMedical Engineering and Informatics,"* 2012. PP 1397 - 1401.

[4] L.Q Sumter, "Cloud Computing: Security Risk," *48th ACM Southeast Regional Conference,* 2010.

[5] Wenchao Zhou et al, "Towards a Data Centric View of Cloud Security," *2nd International Workshop on Cloud Data Management,* 2010, pp 25 – 32.

[6] Ziyuan Wang, "Security and Privacy Issues within the Cloud Computing,"*2011 International Conference on Computational and Information Sciences,* 2011, pp 175 – 178.

[7] John C. Roberts II, Wasim Alhamdani, "Who Can You Trust in the Cloud?: a Review of Security Issues within Cloud Computing," *ACM International Security Curriculum Development Conference,* 2011, pp 15 – 19.

[8] F. Sabahi, "Cloud Computing Security Threats and Responses," *3rd IEEE International Conference on Communication Software and Networks,"* 2011, pp 245 – 249.

[9] Rongxing Liu et al, "Secure Provenance: The Essential Bread and Butter of Data Forensic in Cloud Computing," *5th ACM Symposium on Information, Computer and Communication Security,* 2010, pp 282 – 292.

[10] John Harauz et al, "Data Security in the World of Cloud Computing," *IEEE Computer and Reliability Societies,* 2009, pp 61 – 64.

[11] Kresimir Popvic, Zeljko Hocenski, "Cloud Security Issues and Challenges," 33rd International Convention on Information and Communication Technology, Electronics and Microelectronics, 2010, pp 344 – 349.

[12] Jaeger et al, "Cloud Computing and Information Policy: Computing in a Policy Cloud" *Journal of Information Technology & Politics,* Vol – 5 No – 3.

[13] Dawei Sun et al, "Surveying and Analyzing Security, Privacy and Trust Issues in Cloud Computing Environments", *Procedia Engineering,* Vol. 15, 2011, pp. 2852-2856.

[14] B.R Kandukuri et al, "Cloud Security Issues," IEEE International Conference on Services Computing, 2009, pp 517 – 520.

[15] S.M Hashmi et al, " A Taxonomy of Cloud Computing System: a Servey" International Journal of Applied Information Technology, Vol – 4 No 1, pp 21 – 28.

[16] Neal Levitt, "Is Cloud Computing Ready for the Prime Time?," *IEEE Computer*, Vol – 42, No – 1 pp 15 – 20.

[17] Mladen A. Vouk, "Cloud Computing – Issues, Research and Implementation," Journal of Computer and Information Technology, Vol – 16 No – 4, pp 235 – 246.

[18] D. Zissis, D. Lakkas, " Addressing Cloud Computing Security Issues," Future Generation Computer Systems, Vol – 28, pp 583 – 592.

[19] [12] Craig Gentry, "Computing Arbitrary Functions of Encrypted Data"**,** *Communications of the ACM,* Volume 53 Issue 3, March 2010 pp 97-105

[20] [19] Randike Gajanayake et. al. "Sharing with Care an Information Accountability Perspective," IEEE Internet Computing, 2011, vol. 15, pp. 31-38

[21] [14] Robert Griffin, Subhash Sankuratripati "OASIS Key Management Interoperability Protocol (KMIP)", http://www.oasis-open.org/committees/tc_home.php?wg_abbrev=kmip.

[22] [16] Cong Wang et. al. "Ensuring Data Storage Security in Cloud Computing," *17th* International Workshop on Quality of Service.2009: pp: 1-9.

[23] [15] Zeng K, "Publicly Verifiable Remote Data Integrity," International Conference on Information and Communications Security, 2008, pp: 419 – 434.

[24] Sebstian et. al "Intruder Detection in the Cloud," 8th IEEE International Conference on Dependable, Autonomic and Secure Computing," 2009, pp 729 – 734.